\documentclass[11pt,a4paper]{article}
\usepackage{jinstpub}

\usepackage{amsmath}
\usepackage{graphicx}
\usepackage{xcolor}
\usepackage{natbib}
\usepackage{lineno}

\title{The CaloCube calorimeter for high-energy cosmic-ray measurements in space: performance of a large-scale  prototype}

\author[a,b]{O. Adriani}

\author[c,d]{A. Agnesi}

\author[e,f]{S. Albergo}

\author[g,h]{M. Antonelli}

\author[i,f]{L. Auditore}

\author[l]{A. Basti}

\author[a,b]{E. Berti}

\author[m,l]{G. Bigongiari}

\author[a,b]{L. Bonechi}

\author[a,b]{M. Bongi}

\author[h]{V. Bonvicini}

 \author[b]{S. Bottai}

\author[m,l]{P. Brogi}

\author[n,b]{G. Castellini}

\author[d]{P.W. Cattaneo}

\author[o,l]{C. Checchia}

\author[a,b]{R. D'Alessandro}

\author[b]{S. Detti}

\author[p,q]{M. Fasoli}

\author[r,b]{N. Finetti}

\author[f]{A. Italiano}

\author[m,l]{P. Maestro}

\author[m,l]{P.S. Marrocchesi}

\author[a,b]{N. Mori}

\author[h]{G. Orzan}

\author[a,b]{M. Olmi}

\author[a,b,n]{L. Pacini}

\author[b]{P. Papini}

\author[f]{M.G. Pellegriti}%

\author[c,d]{F. Pirzio}%

\author[h]{C. Pizzolotto}

\author[a]{C. Poggiali}%

\author[d]{A. Rappoldi}

\author[n,b]{S. Ricciarini}

\author[s]{A. Sciuto}

\author[a]{P. Spillantini}

\author[b]{O. Starodubtsev}

\author[m,l]{F. Stolzi}

\author[m,l]{J.E. Suh}

\author[m,l]{A. Sulaj}

\author[a,b]{A. Tiberio}

\author[e,f]{A. Tricomi}

\author[i,f]{A.Trifiro}

\author[i,f]{M. Trimarchi}

\author[p,q]{A. Vedda}

\author[b,1]{E. Vannuccini \note{Corresponding author.}}

\author[h]{G. Zampa}
\author[h]{N. Zampa}

\affiliation[a]{Department of Physics and Astronomy, University of Florence, \\
via G. Sansone 1, I-50019 Sesto Fiorentino (Firenze), Italy}
\affiliation [b]{INFN Firenze, \\
via B. Rossi 1, I-50019 Sesto Fiorentino (Firenze), Italy}
\affiliation [c]{Dipartimento di Ingegneria Industriale e dell'Informazione, University of Pavia,\\
Pavia, Italy}
\affiliation [d]{INFN Pavia, \\
via A. Bassi 6, I-27100 Pavia, Italy}
\affiliation [e]{Department of Physics and Astronomy, University of Catania, \\
via S. Sofia 64, I-95123 Catania, Italy}
\affiliation [f]{INFN Catania, \\
via S. Sofia 64, I-95123 Catania, Italy}
\affiliation [i]{Dipartimento di Scienze Matematiche e Informatiche, Scienze Fisiche e Scienze della Terra, University of Messina,  \\
sal. Sperone 31, I-98166 Messina, Italy}
\affiliation [m]{Department of Physical Sciences, Earth and Environment, University of Siena, \\
I-53100 Siena, Italy}
\affiliation [l]{INFN Pisa, \\
Largo B. Pontecorvo, 3 – 56127 Pisa, Italy}
\affiliation [h]{INFN Trieste, \\
via Valerio 2, I-34127 Trieste, Italy}
\affiliation [n]{IFAC (CNR), \\
via Madonna del Piano 10, I-50019 Sesto Fiorentino (Firenze), Italy}
\affiliation [o]{Department of Physics, University of Pisa, \\
Largo Bruno Pontecorvo 3, 56127 Pisa}
\affiliation [p]{Department of Materials Science, University of Milano-Bicocca, \\
via Cozzi 55, I-20125 Milano, Italy}
\affiliation [q]{INFN Milano-Bicocca, \\
Piazza della Scienza 3, Milano, Italy}
\affiliation [r]{Department of Physical and Chemical Sciences, University of L'Aquila, \\
Via Vetoio, Coppito, 67100 L'Aquila, Italy}
\affiliation[g]{Department of Physics, University of Trieste, \\
via Valerio 2, I-34127 Trieste, Italy}
\affiliation[s]{MATIS (CNR), \\
via S. Sofia 64, I-95123 Catania, Italy}

\emailAdd{vannuccini@fi.infn.it}

\abstract{
The direct observation of high-energy cosmic rays, up to the PeV energy region, will increasingly rely on highly performing calorimeters, and the physics performance will be primarily determined by their geometrical acceptance and energy resolution. 
Thus, it is extremely important to optimize their geometrical design, granularity and absorption depth, with respect to the total mass of the apparatus, which is amongst the most important constraints for a space mission. 
CaloCube is an homogeneous calorimeter whose basic geometry is cubic and isotropic, obtained by filling the cubic volume with small cubic scintillating crystals. In this way it is possible to detect particles arriving from every direction in space, thus maximizing the acceptance.

This design summarizes a three-year R$\&$D activity, aiming to  both optimize and study the full-scale performance of the calorimeter, in the perspective of a cosmic-ray space mission, and investigate a viable technical design by means of the   construction of several sizable prototypes. 
A large scale prototype, made of a mesh of $5 \times 5 \times 18 $ CsI(Tl) crystals, has been constructed and tested on high-energy particle beams at CERN SPS accelerator.
In this paper we describe the CaloCube design and present the results relative to the response of the large scale prototype to electrons.
}


\begin{document}

\maketitle

\flushbottom
\section{Introduction}

Indirect measurements, performed by detecting on ground the extensive air showers produced by primary Cosmic Rays (CRs) in the atmosphere, show that around the PeV energy region the inclusive spectrum of particles becomes suddenly steeper and the composition progressively heavier. 
This feature, known as the CR ``knee", is believed to indicate the energetic limit of the galactic accelerators. 
A precise knowledge of particle spectra and composition in this spectral region would allow to address key items in the field of high-energy CR physics, such as the unambiguous identification of the acceleration sites, the clear understanding of the acceleration mechanisms, as well as an accurate modeling of particle propagation and confinement within the Galaxy.  
In spite of the improvements achieved by indirect techniques, composition studies are still very difficult. 
Only the spectra of groups of elements are measured and the results are considerably model dependent, concerning both the energy reconstruction and the element identification (see e.g. \cite{kascade,argo}).

Direct CR detection permits unambiguous elemental identification and a more precise energy measurement.
The typical instrumental configuration used to perform the highest-energy direct measurements of the CR elemental spectra is a calorimeter coupled to a charge measuring device. 
This setup has proven very powerful to study the high energy cosmic radiation in space, since it can be easily scaled to cover the desired energy range. 
Recently, the CALET~\cite{PhysRevLett.122.181102} and DAMPE~\cite{An:2019wcw,PhysRevLett.126.201102} experiments reported results on the H and He spectra over a wide energy range,  covering the disjunct sub-ranges explored by previous calorimetric instruments (ATIC-2~\cite{Panov:2011ak}, CREAM~\cite{Yoon:2017qjx,Yoon:2011aa}, and NUCLEON~\cite{Atkin2018}) and magnetic spectrometers (PAMELA~\cite{Adriani:2017bfx}, and AMS-02~\cite{PhysRevLett.114.171103,PhysRevLett.115.211101}), and revealed interesting spectral features. 
However, space and balloon-borne experiments suffer from limitations on the effective acceptance that can be achieved, which  practically prevents the present missions to go beyond 100 TeV,  due to the steepness of the CR spectra. 
Even more severe limitations affect the less-abundant heavier nuclei, and in particular the rare secondary boron component, whose abundance provides the most stringent constraint to propagation models and is measured only up to $\approx$~1~TeV/n \cite{PhysRevLett.120.021101}. 

Calorimetric CR measurements conducted in space have in their reach also the capability to study the inclusive electron component (electrons+positrons) and, if the calorimeter is coupled to a tracker-converter system and an anti-coincidence shield, the high-energy gamma radiation. 
The CALET and DAMPE experiments have been specifically designed to investigate the high-energy electron spectrum, with the purpose of resolving the conflicts among previous measurements (AMS-02~\cite{PhysRevLett.113.121102}, Fermi/LAT~\cite{PhysRevD.95.082007}, HESS~\cite{PhysRevLett.101.261104,Aharonian:2009ah}) and extending the direct observation above 1~TeV, where a cut-off of the diffuse galactic component is expected, and indirectly observed by HESS, and possible contributions from nearby sources could emerge. 
They both carried out measurements extending up to 4.8~TeV~\cite{Ambrosi2017,PhysRevLett.120.261102}  finding however contradicting results, which indicates, 
in spite of the very good energy resolution of the calorimeters, the presence of still unknown systematic errors. 

The latest experimental efforts motivated an extensive R$\&$D activity aiming to further improve the  performance of calorimetric measurement in space.   
In order to clearly detect the ``knees'' of the individual H and He spectra, an acceptance of at least  2.5 m$^2$sr~$\times$~5~yr is necessary and an energy resolution better than 40$\%$ is desirable.  
In order to improve the quality of the measurements on the electron+positron flux, an excellent energy resolution for electromagnetic showers 
and a high electron/hadron rejection power 
are required, which pose further constraints on the instrument. 

CaloCube \cite{CC0,CC1,bongi2015calocube,vannuccini2017calocube,adriani2016calocube,adriani2017calocube,pacini2017calocube,berti2019calocube, mc,Adriani2019} is an R$\&$D project  aiming to optimize the design of a wide-acceptance 3-D imaging calorimeter to be operated in space~(section~\ref{sec:design}). 
The CaloCube concept design has been adopted by the future High Energy cosmic-Radiation Detection (HERD) experiment~\cite{Dong:2020ugg}, that has been proposed as one of several space astronomy payloads on board the future China’s Space Station. 
HERD is conceived as a multipurpose science facility, that will perform indirect dark matter search, cosmic-ray spectrum and composition measurements up to the ``knee'' energy, as well as  
gamma-ray monitoring and full sky survey, during a 10-year mission that is planned to start the operations around 2025.   
The core of the HERD apparatus is a deep 3-D  imaging calorimeter, made of LYSO crystals, which partly implements the CaloCube design. 


\subsection{CaloCube concept and overall design }
\label{sec:design}
To achieve the  performance required to extend the range of direct CR nuclei measurements up to the PeV region with a space-borne calorimeter is definitely a challenge.
The major constraint comes from the limitation in mass for the apparatus (few tons), which severely affects both the geometrical factor and the energy resolution.

\begin{figure}[t]
\begin{center}
\includegraphics[width=15cm]{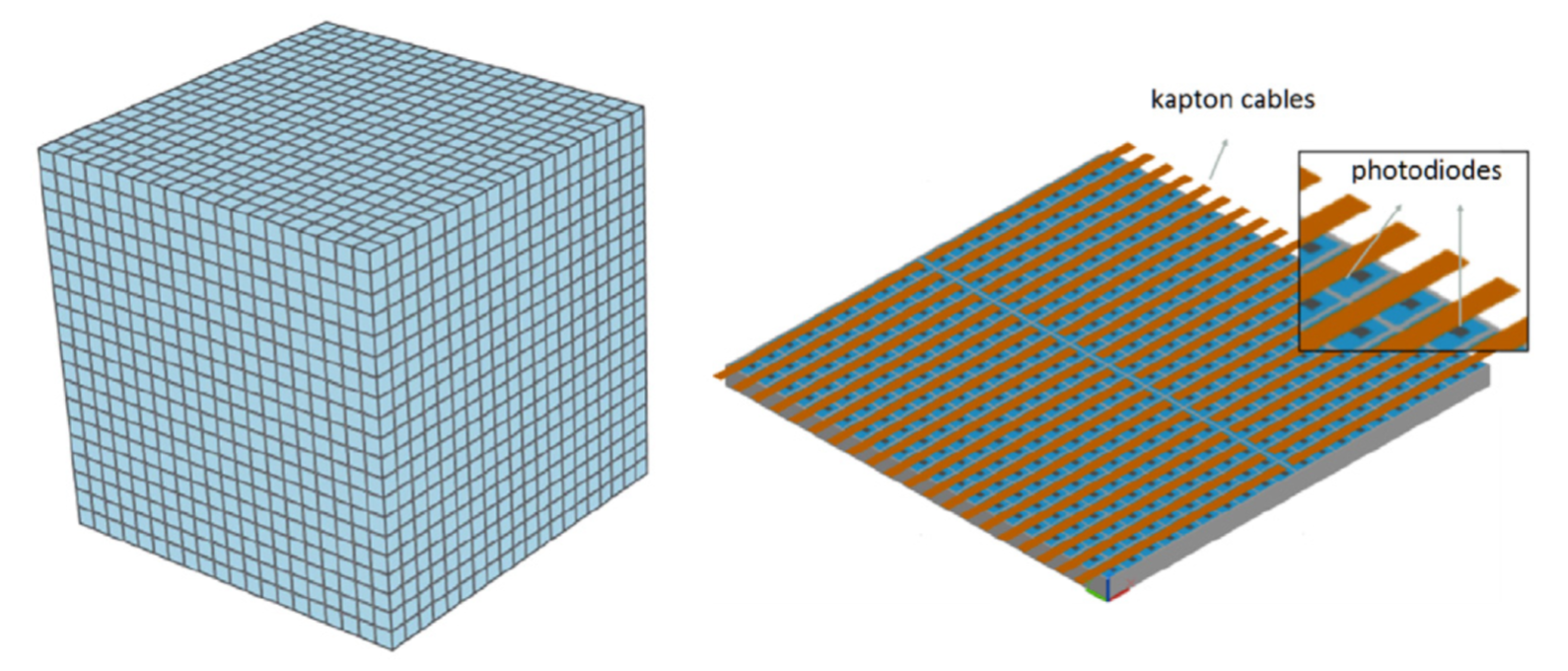}
\end{center}
\caption{ Basic design of the CaloCube detector. The  drawing on the left shows the entire active volume, made of $\approx$~2~tons of CsI(Tl), segmented into 21$\times$21$\times$21 cubic crystals of side equal to 3.6~cm (about 1 Moli\`ere radius). 
The drawing of the right shows one plane of cubic crystals with the dedicated photodiodes and kapton cables for the read-out of the light signals.  \label{fig:cc}}
\end{figure}

The proposed solution consists in a 3D array of cubic scintillating crystals, read out by photodiodes (PDs), arranged to form a cube (see Figure~\ref{fig:cc}). 
The cubic geometry and the homogeneity provides the possibility to collect particles from either the top and the lateral facets, thus allowing to maximize the geometrical acceptance for a fixed mass budget. 
The active absorber provides good energy resolution, while  
the high granularity allows to perform shower imaging, thus providing criteria for both leakage correction and electron/hadron separation. 

An extensive Monte Carlo simulation study, based on the FLUKA~\cite{flukaa,flukab} package, was carried out, focusing on the optimization of the CaloCube conceptual design for nuclei detection in the TeV$\div$PeV energy range~\cite{mc}.  
A total weight of 2 tons was assumed, including both active and passive materials. 
A comparative study of suitable scintillating materials was performed, spanning a wide range of densities, calorimetric and optical properties. 
An effective geometric factor up to about 4~m$^2$sr can be achieved, with an energy resolution better than 40$\%$~\cite{mc}.  
In particular, the results favor materials which provide better shower containment (e.g. LYSO), which compensates for the smaller volume due to the larger density and smaller interaction length typical of these crystals.

From the point of view of electrons and gamma rays,  
the presence of passive materials among the scintillating crystals and the direct ionization of the PDs introduce fluctuations in the collected energy that might be significant. 
A preliminary study, performed simulating a 2~tons calorimeter made of CsI(Tl), with crystals of 3.6~cm size and 0.4~cm spacing, indicated that a 2$\%$ energy resolution can be achieved  with an isotropic flux of electrons of energy between 100~GeV and 1~TeV~\cite{CC1}.

This paper focuses on the performance of a large-scale prototype made of CsI(Tl) crystals~(section~\ref{sec:prototype}),  that has been constructed and tested with high-energy particle beams at the CERN SPS. The instrument has been calibrated using beam-test data  (section~\ref{sec:calibration}) and the response to electromagnetic showers has been studied, comparing simulated (section~\ref{sec:simulation}) and real data (section~\ref{sec:performance}).

\section{Calorimeter prototype }
\label{sec:prototype}

As a proof of principle of the CaloCube concept, several prototypes of the calorimeter,  of progressively increasing size, have been constructed in recent years. 
Many tests have been performed in laboratory, in order to optimize the light collection efficiency, to compare and characterize various PD responses and to test system readout. 
An extensive review of these efforts is presented in reference~\cite{Adriani2019}.

\begin{figure}[t]
\begin{center}
\includegraphics[width=7.5cm]{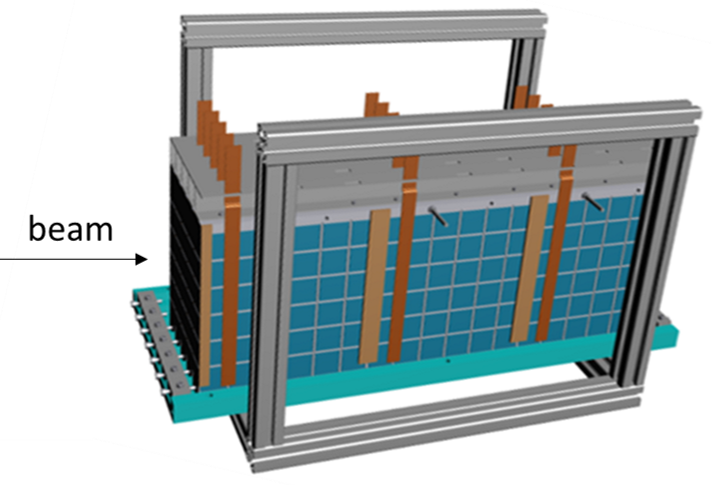}
\includegraphics[width=7.5cm]{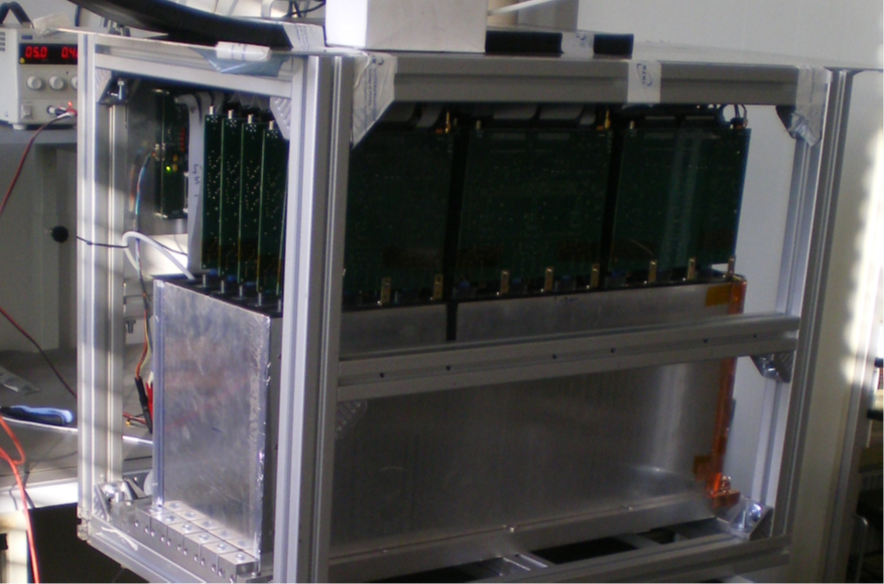}
\end{center}
\caption{Mechanical design  of the fully assembled prototype (left) and its realization (right). In the mechanical design only one, out of the three, kapton cable for each tray is shown. Five trays are stacked to form a module. Three modules are placed side by side, for a total length of 18 crystals along the beam direction. 
The whole blocks were enclosed within an aluminum box (right). \label{fig:pro}}
\end{figure}

A picture of the largest prototype constructed  so far is shown in Figure~\ref{fig:pro}.  
This calorimeter prototype 
is composed by a total of 30 Polyoxymethylene trays,
each having a matrix of 6$\times$6 squared cavities, hosting the scintillating crystals, placed 4 cm apart from each other.
Five trays are stacked to form a module. 
Three modules are placed side by side, for a total length of up to 18 crystals along the beam direction.

\begin{figure}[t]
\begin{center}
\includegraphics[width=8.5cm]{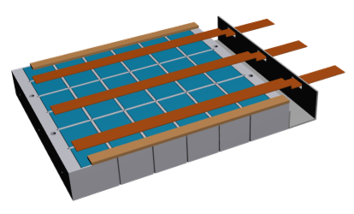}
\includegraphics[width=6.5cm,angle=90]{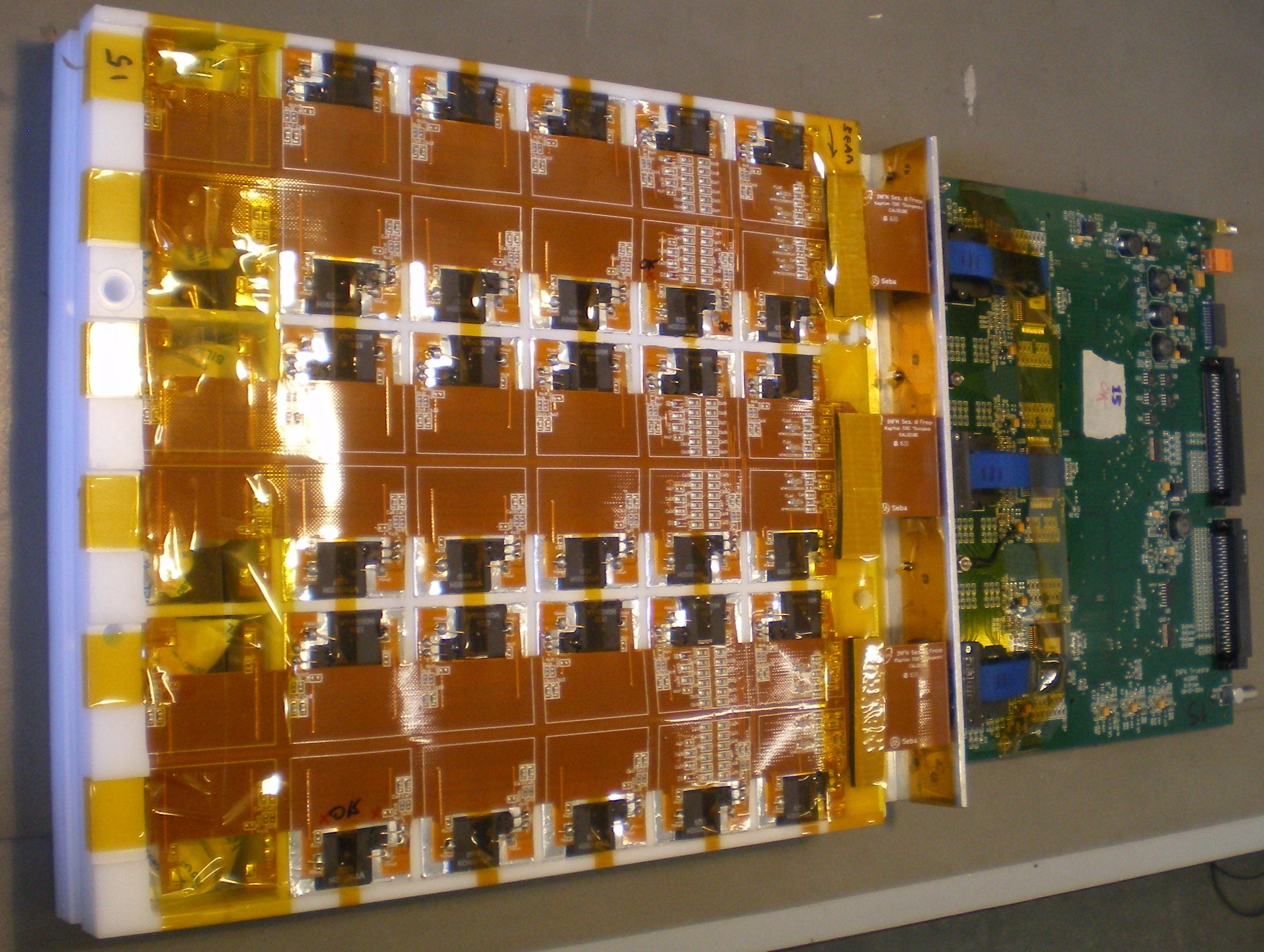}
\end{center}
\caption{Mechanical design of a Polyoxymethylene tray equipped with CsI (Tl) crystals (left) and its realization (right).   Also visible in the picture (right) are the photodiodes, with the kapton cables used to route the  signals to the front-end  board. 
The bottom row of the tray was not used in the prototype.
\label{fig:mec}}
\end{figure}

A picture of one tray is shown in  Figure~\ref{fig:mec}. 
Crystals are inserted into the tray cavities with the PDs placed on the external facet. 
Signals coming from the PDs are readout by means of kapton flexible printed circuit boards and brought to the front-end board, which is placed on the side of each tray. 

For the beam test, each tray was equipped with a matrix of 6$\times$5 3.6~cm size crystals (see the right picture in figure~\ref{fig:mec}) so that the full prototype was hosting a total of 450 crystals arranged in a 5$\times$5$\times$18 matrix.

\subsection{Crystals and photodiodes} 

\begin{figure}[t]
\begin{center}
\includegraphics[width=7cm]{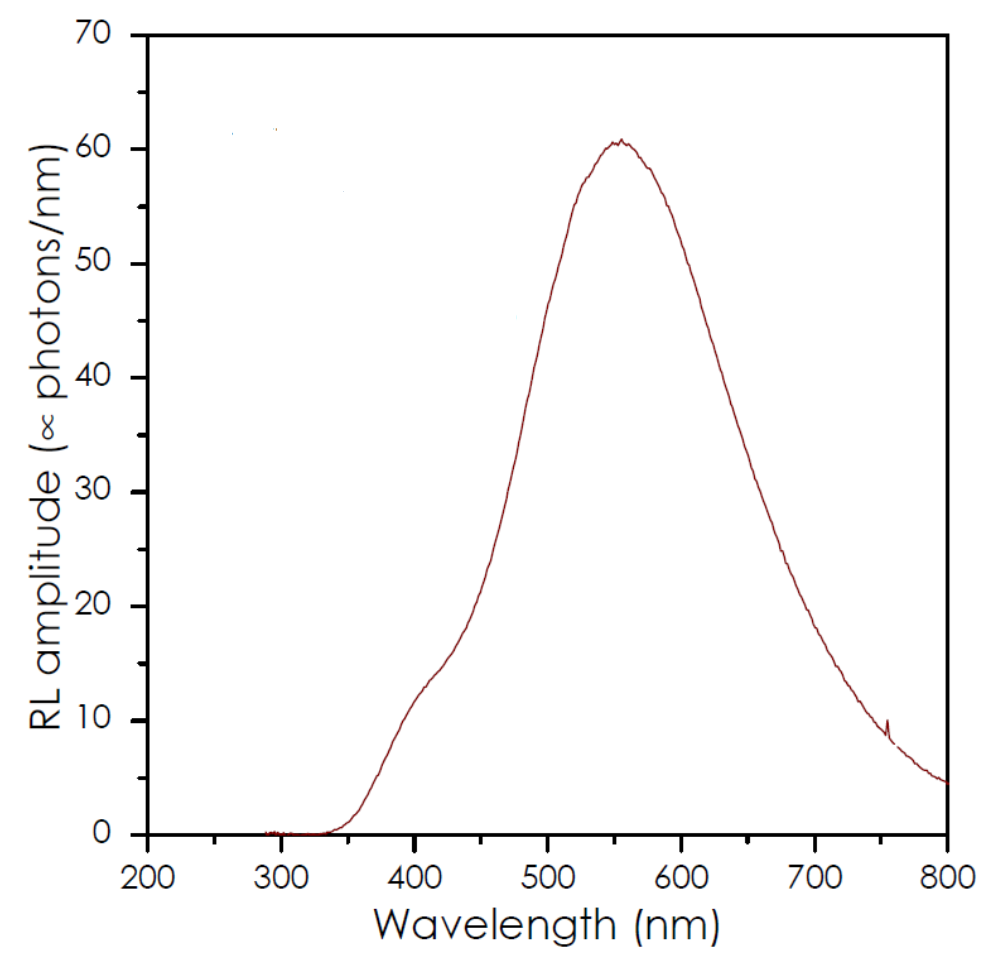}
\end{center}
\caption{ X-ray induced radioluminescence (RL) spectrum of one of the CsI(Tl) crystals used for the prototype. \label{fig:csitl}}
\end{figure}

The scintillating material chosen for the Calocube prototype is CsI(Tl). 
This choice has been dictated mainly by practical reasons; CsI(Tl) is widely available on the market at an affordable price, 
it has a very high light yield and  its emission spectrum matches very well the spectral response of a large variety of Si PDs.  
In spite of its lower calorimetric performances, if compared with other inorganic crystals like BGO or LYSO~\cite{mc}, CsI(Tl) has allowed to construct, at reduced cost, a large scale prototype with good containment capabilities, apt to demonstrate the validity of the CaloCube concept. 
The resulting  total depth of the prototype, along the longest dimension, is $\approx$~1.3 nuclear interaction lengths and $\approx$~27 radiation lengths.

The crystals used for the prototype have been produced by AMCRYS\footnote{Web: http://www.amcrys.com/ .}. 
Figure~\ref{fig:csitl} shows their typical emission spectrum, which spans over a wide range of wavelengths and peaks at about 550~nm. 
Each crystal has been wrapped in a reflective Vikuiti$^{TM}$ film, which has resulted the best choice to maximize the light collection efficiency~\cite{Adriani2019}.\footnote{Web: www.3m.com/uk/vikuiti .}

\begin{table}[]
\begin{center}
\begin{tabular}{l|ll}
                                 & VTH2090H       & VTP9412H \\ \hline
Active area (mm$^2$)   & 84.6               &  1.6 \\
C$_J$ (pF)                   & 70 (@30V)      & 6 (@15V)  \\
Spectral response range  (nm)    & 400$\div$1100 & 400$\div$1150 \\
Quantum efficiency (@550 nm)              & 69$\%$          & 75$\%$
\end{tabular}
\caption{ Summary of the main characteristics of the photodiodes used for the calorimeter prototype. \label{tab:pds}}
\end{center}
\end{table}

One of the most challenging requirement for the instrument is the very large dynamic range needed to detect PeV protons. 
According to simulation~\cite{mc}, an interacting proton can deposits up to 10$\%$ of its kinetic energy in a single CsI(Tl) crystal of 3.6 cm size.
Considering that not-interacting minimum ionizing protons deposits  $\approx$~20 MeV, the needed dynamic range is of the order of 10$^7$. 
This will be accomplished by using multiple PDs of different active area: a Large area (LPD) and a Small area PD (SPD). In such a way, when the signal in the LPD is so large to induce the saturation of the electronics, the energy deposit in that crystal can be measured exploiting the information from the SPD. 

For the prototype, each crystal is coupled to two different commercially available PDs, having similar spectral response but different area:  
the VTH2090H and the VTP9412H, both produced by Excelitas\footnote{Web: https://www.excelitas.com/ .}.  
Their main characteristics are summarized in Table~\ref{tab:pds}. 
The PDs are optically coupled to the crystals with silicone glue and their signals are routed to the front-end (FE) board by means of flexible kapton cables, visible on the left picture of figure~\ref{fig:pro}.

The present double-PD design does not yet provide the dynamic range  required by the cosmic-ray mission. 
Several solutions exist to increase the dynamic range, e.g. by applying an optical filter in front of the SPD.

\subsection{Read-out electronics} 
\label{sec:chip}

\begin{figure}[t]
\begin{center}
\includegraphics[width=8cm]{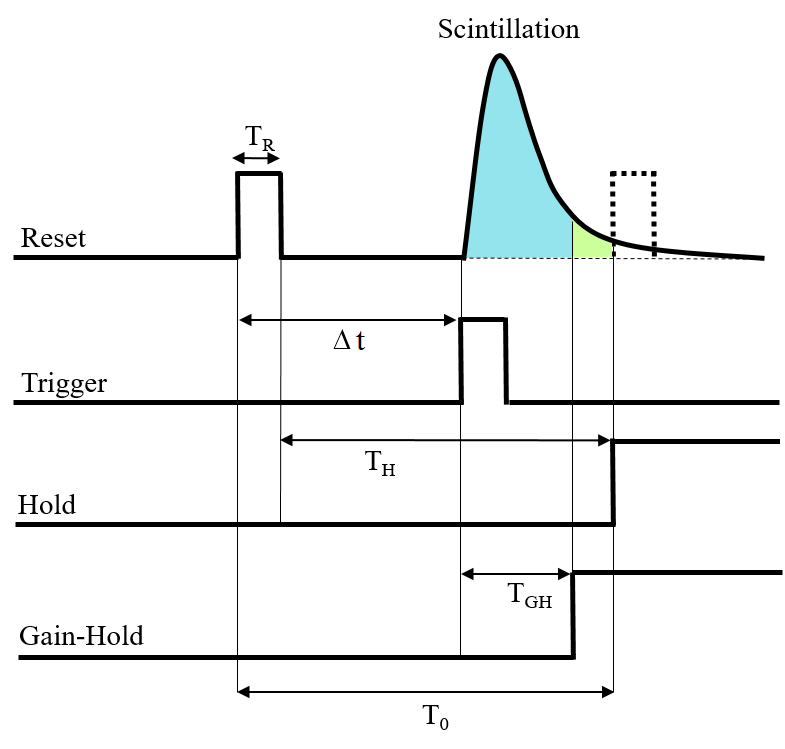}
\end{center}
\caption{ Simplified timing sequence of the FE chip. When a {\it Trigger} signal is sent to the chip, the {\it Reset} is inhibited and two signals, {\it Gain-Hold} and {\it Hold}, are generated  after a fixed adjustable delay, $T_{GH}$ and $T_{H}$, to lock the automatic gain and the sampled voltage, respectively;  
while the {\it Hold} is synchronous with the {\it Reset}, the {\it Gain-Hold} is synchronous with the {\it Trigger}. For the prototype $T_{H}$~=~20~$\mu$s, $T_{GH}$~=~7~$\mu$s, $T_{R}$~=~0.7~$\mu$s.
\label{fig:time}}
\end{figure}

The FE electronics is based on a high dynamic-range,  low-noise ASIC, developed by members of the CaloCube collaboration and specifically designed for Si-calorimetry in space \cite{CASIS,Adriani2019}.   
The wide gain of the chip is obtained through a double-gain charge-sensitive amplifier, where the gain is selected in real time depending on the amount of integrated charge. 
When the system switches to low gain, the chip linear range is expanded by a factor $G  \approx$~20. 

A relevant feature of the chip, when used with scintillating crystals, is that it must be reset periodically. 
A simplified scheme of the time sequence of the chip is illustrated in figure~\ref{fig:time};  
when a {\it Trigger} signal is sent to the chip, the {\it Reset}  is inhibited 
and two signals, {\it Gain-Hold} and {\it Hold}, are generated to lock the automatic gain and the sampled voltage, respectively. 
While the {\it Hold}  
replaces the subsequent {\it Reset}, 
the {\it Gain-Hold} is synchronous with the {\it Trigger}.  
For the calorimeter prototype the {\it Hold} delay  $T_{H}$ has been set to 20~$\mu$s and a long {\it Gain-Hold} delay $T_{GH}$ of 7~$\mu $s has been set in order to assure that most of the CsI(Tl) scintillating signal is integrated before locking the gain, thus avoiding signal saturation in high-gain.  
The chip is considered to behave nominally when $\Delta t$ is within the range $T_R \div (T_R+T_H-T_{GH}) $, where $T_R$ is the duration of the {\it Reset} signal. 

This chip architecture offers many advantages for the design of a calorimeter to be operated in space, e.g. a wide dynamic range with a low power consumption. 
However, it poses some challenges when used to readout slow signals, typical of the CsI(Tl) scintillation light response, since for instance the integration time is variable.   
In order to monitor the chip behaviour, the time distance $\Delta t$ between the rising edge of {\it Trigger} and the rising edge of the last  {\it Reset} is readout for each event. 
In this work the data analysis is restricted to the time range $\Delta t$~=~0.7$\div$14~$\mu$s, where the chip is active and the gain is locked. 

The chip version used for the calorimeter prototype is HIDRA v1.0, that has 28 channels.
One chip is used to readout up to 12 crystals (two lines for each tray), which require a total of 24 channels. 
Two of the remaining channels are connected to diodes placed on the two sides of the kapton cable  
and are used to monitor the common noise; 
these channels are from now on referred to as CN channels.
The other channels were not used.

\subsection{Test beams}

The calorimeter prototype has been exposed to high-energy particles at the secondary H4 beam line of the Super Proton Synchrotron (SPS) at CERN, during two separated test sessions, in 2017.

 The first test was carried out with secondary particle beams of different types and momenta: 50~GeV/c muons, 50$\div$280~GeV/c electrons and  100$\div$350~GeV/c mixed hadrons. 
 A micro-strip Si tracking system~\cite{ADAMO} was placed upstream the prototype and provided the particle impact position with a resolution of $\approx$~1~mm.   
 The second test was carried out  with ions produced by the fragmentation of $^{129}$Xe primary beam of 150~GeV/A kinetic energy hitting a 40~mm Be target.

In this paper we focus on the results obtained from the first beam-test data, that were used to   calibrate the detector and study the performances relative to the electromagnetic showers. 
Results on the response to hadronic showers will be the subject of a forthcoming paper.

\section{ Calorimeter calibration }
\label{sec:calibration}

In the first part of this section we describe the single-channel output and the system noise performance  (section~\ref{single:noise}).
The second part of the section is dedicated to the instrument calibration.  
Signals from each channel were  equalized  to account for the different responses of the PDs coupled to different CsI(Tl) crystals. 
The gain factors for LPDs and SPDs were determined from the beam data, as discussed in Sections~\ref{single:mip} and~\ref{sec:spdgain}, respectively.

\subsection{Single-channel response}

\label{single:noise}

\begin{figure}[t]
\begin{center}
\includegraphics[width=7.5cm]{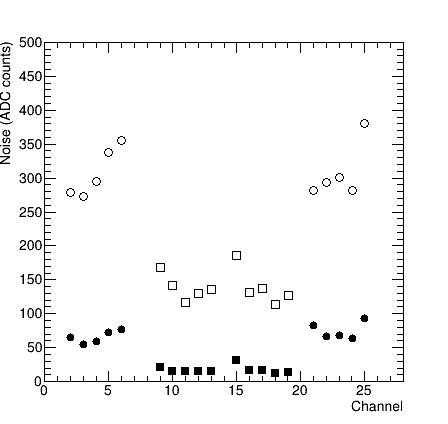}
\includegraphics[width=7.5cm]{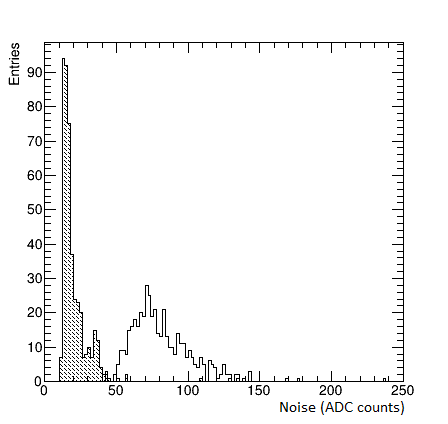}
\end{center}
\caption{Left: noise, evaluated as the RMS of the signal readout from the channels of one of the FE boards, before (empty markers) and after (solid markers) common noise subtraction. Cyrcles (squares) refer to channel connected to large (small) PDs. Right: Distribution of channel noise for large PDs (empty area) and small PDs (dashed area). \label{fig:noise}}
\end{figure}

Noise and pedestals of the 900 readout channels (see section~\ref{sec:prototype}) were  monitored during the beam test by acquiring off-spill events. 
Figure~\ref{fig:noise} (empty markers) shows the typical RMS of the signals readout from channels connected to one of the HIDRA chips.
Only the channels connected to crystals, 20 channels out of 28, are shown. 
This version of the prototype was characterized by a quite large noise. 
This behaviour has been ascribed to the flexible connection circuits that bring the PDs signal to the readout electronic boards, that were not fully optimized for this prototype. A lower noise is observed for channels connected to the SPDs (square markers), due to the lower capacitance of the sensors with respect to the LPDs (circle markers).

\begin{figure}[t]
\begin{center}
\includegraphics[width=7.5cm]{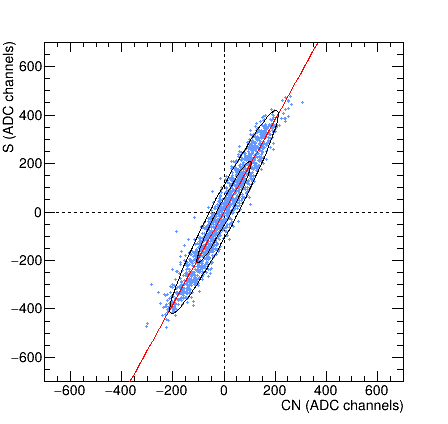}
\includegraphics[width=7.5cm]{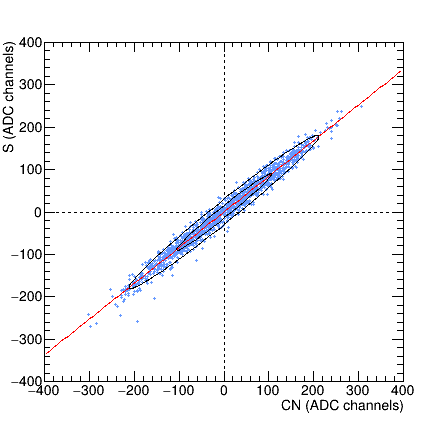}
\end{center}
\caption{ Distribution of the signal (S) readout from the large (left) and small (right) PD channels  of one of the CsI(Tl) crystals as a function of the signal (CN) readout from the test diode placed on the same side of the kapton cable, for a sample of off-spill events.   \label{fig:cn}}
\end{figure}

A detailed analysis of the system noise was performed by studying the correlation  among signals readout from channels connected to PDs and signals readout from channels connected to the test diodes placed on the kapton cables (CN).
The largest correlation coefficient has been observed for channels connected to readout lines placed on the same side of a kapton cable. 
A lower correlation coefficient was observed between the two sides of the cable, while channels connected to different cables can be considered uncorrelated. 
As an example, figure~\ref{fig:cn} shows the variation of the two PD signals, a large and a small one, as a function of the CN signal readout from the same side of the kapton cable, after pedestal (PED) subtraction.  
The scatter plots in the pictures highlight a strong linear correlation between the PD signals and the CN signal, indicating that a large fraction of the noise fluctuations is common to all channels. 

On the basis of these considerations, the common noise component for each PD channel was evaluated event-by-event assuming a linear relation with the CN signal registered on the same side of the kapton cable.
The signal of the $i^{th}$ PD channel, placed respectively on the $a,b$ side of a kapton cable,  was evaluated for the $n^{th}$ event as:  
\begin{equation}
S_{i,n} = R_{i,n} - PED_{i} - \alpha_{i}^{a,b} \cdot CN^{a,b}_n
\label{eq:pedcnsub}
\end{equation}
where $R_{i,n}$ is the raw value of the signal readout from that channel and $PED_{i}$ is the pedestal. 
The parameters $PED_{i}$ and $\alpha_{i}^{a,b}$ were evaluated, for each particle-beam run, from the sample of acquired off-spill events:
the pedestal levels were determined as $PED_{i}=\langle R_{i} \rangle$ and was  $\approx$~17900~ADC~counts;  
the coefficients  $\alpha_{i}^{a,b}$ were  determined from the slopes of the correlation ellipses. 
The residual noise, evaluated as the root mean square (RMS) of  $S_{i}$, for the sample of off-spill events, is shown in figure~\ref{fig:noise} (left). 
The signal fluctuations is significantly reduced after common noise subtraction, and the average residual noise for LPDs and SPDs amount to about 85 and 20 ADC counts figure~\ref{fig:noise}~(right).

As discussed in section~\ref{sec:chip}, when the signal integrated over the time interval $T_{GH}$ exceeds the threshold value ($\approx$~46000~ADC~counts) the gain is  automatically reduced by a factor $1/G \approx 1/20$; $1/G$  is determined by the size of the feedback loop capacitors.

%
When a signal $R_{i,n}$ is readout in low-gain regime, after pedestal and common noise  subtraction, it was scaled according the relation:
\begin{equation}
S_{i,n} = G \cdot (R_{i,n}-PED_i- c \cdot \alpha_i^{a,b} \cdot CN_n^{a,b}) + P  
\label{eq:lg}
\end{equation}
The parameters in Eq.\ref{eq:lg} represent the ratio ($G$) between high and low gain of the HIDRA chips, a pedestal offset ($P$) and a noise scaling factor ($c$).  
The scaling parameters were measured in laboratory during the qualification test of the HIDRA board prototypes. The high-to-low gain ratio was found to agree with the nominal value $G=20$ at the level of less than $0.5\%$. The pedestal shift $P$ was found to be not negligible and equal to $\approx -2000$~ADC~units. 
The factor $c$ accounts for the fact that
in low-gain mode the noise is reduced by a factor $\approx$~0.07 and $\approx$~0.11 for LPD and SPD channels, respectively. 
All three parameters were assumed to be the same for all the HIDRA boards. 
The low-gain single-channel response is linear up to a threshold value that depends on the board and is  $\gtrapprox$~51000~ADC~counts; above this value the signal sharply saturates.

Due to capacitive coupling, in case of saturation of a channel, a fraction of the charge is injected into nearby channels. The amount of charge injection is quantified using the CN channel: a small drift of the CN signal was indeed observed, of the order of $\approx$~10$^{-4}$ times the maximum signal readout from the same side of the kapton cable. 
In case of strong saturation of a single channel, due to large signal in the corresponding LPD, the CN signal was found to sharply increase.
The effect was studied in laboratory by inducing saturation of a single  channel connected to a LPD: 
the test revealed a significant signal injection, proportional to the CN signal, into the nearby channels, including the readout channel connected to the SPD coupled to the same crystal of the saturated LPD channel. 
For each channel we estimated the correlation between the charge injected into the CN channel and the channel itself. For most channels, these correlation factors resulted to be very similar to the one estimated for common noise subtraction. This was not true for the channel connected to the SPD coupled to the same crystal of the saturated LPD channel, which in this case has a correlation factor of $\alpha \approx  3$ (see Eq.~\ref{eq:pedcnsub} and \ref{eq:lg}). By acting in this way, we were able to partially compensate the charge injected from the saturated channel.

\subsection{LPD gain calibration}
\label{single:mip}

The most probable value of the energy deposit of Minimum Ionizing Particles (MIPs) in a single CsI crystal is about 20~MeV in the vertical direction and represents the minimum signal the crystals should be able to detect.
MIP signals are detected in the high-gain regime of the LPD. 
The response of single crystals to MIPs was studied with the muon beam at 50~GeV. 

After common-noise subtraction, evaluated as described in section~\ref{single:noise}, the MIP signal is well separated from the residual noise, but it manifests a significant time dependence. 
This feature is related to the working principle of HIDRA chip, illustrated in section~\ref{sec:chip}, joined to the slow light-response typical of the CsI(Tl) scintillating crystal.
Since the emission time of the scintillation light in CsI(Tl) is comparable with the hold-signal delay (T$_H$ = 20~$\mu$s for the HIDRA setup used for this work),  when a trigger signal is generated close to the end of the live-time interval 
the integration time is insufficient to collect the whole scintillation signal.   

The responses of the LPDs were equalized by correcting  the acquired signals  for both the gain dispersion among different crystals and for the time   attenuation. 
For this purpose, the single crystal response was  expressed as $S(\Delta t ) = S_0 \cdot a(\Delta t )$. 
While  the average value of $S_0$ for MIPs varied significantly from crystal to crystal, a uniform attenuation function, $a(\Delta t)$, was assumed for the whole calorimeter.  

\subsubsection{Timing correction}
\label{single:time}

The time variation of the CsI(Tl) scintillation signal is well reproduced by assuming a negligible rise time and two decay components, a fast and a slow one, characterized by the decay times $\tau_{f}$ and $\tau_{s}$ respectively. 
In this approximation the time evolution of the signal can be expressed as:
\begin{equation}
I(t) = I_{0} [f \cdot e^{-t/\tau_f}+(1-f)\cdot e^{-t/\tau_s}]
\end{equation}
where $f$ is the fast emission yield fraction and  $I_{0}$ the absolute intensity of the signal.
Given the timing sequence described in section~\ref{sec:chip} (figure~\ref{fig:time}), the signal integrated by the circuit as a function of the time interval $\Delta t$ can be expressed as:
\begin{equation}
\begin{aligned}
S(\Delta t )  & = 
 \int _{0} ^{T_0-\Delta t} I(t) dt \\
  & = 
 I_{0} [ f  \tau_f  (1-e^{-T_0/\tau_f} \cdot e^{\Delta t/\tau_f}) 
  +  (1-f)  \tau_s  (1-e^{-T_0/\tau_s} \cdot e^{\Delta t/\tau_s} ) ]
 \end{aligned}
\label{eq:timecorr}
\end{equation}

\begin{figure}[t]
\begin{center}
\includegraphics[width=7.5cm]{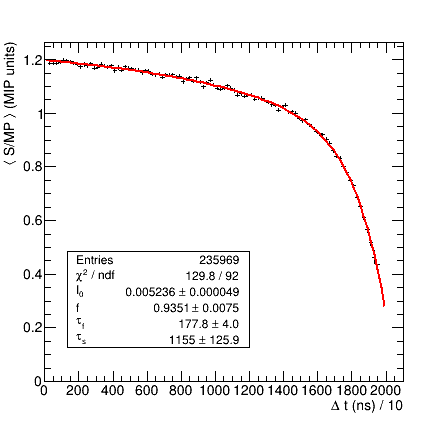}
\includegraphics[width=7.5cm]{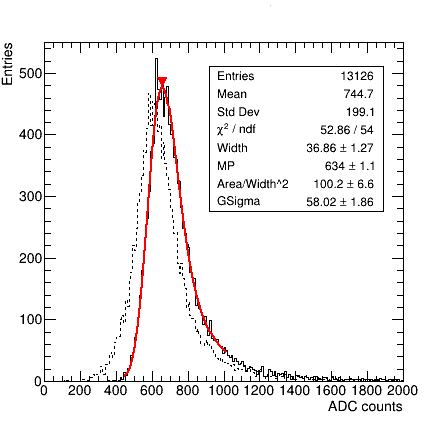}

\end{center}
\caption{Left: average signal induced by 50~GeV muons on single crystals, divided by the crystal gain, as a function of the time difference ($\Delta t$) between the {\it Reset} and the {\it Trigger} signals (see the timing sequence in figure~\ref{fig:time}). The red curve shows a fit to the data with a function defined by Eq.~\eqref{eq:timecorr}, which is used to correct event-by-event the signals for the time attenuation. Right: distribution of the signal for one of the crystals, before (dashed line) and after (solid line) the timing correction. The red curve shows a fit of the distribution with the convolution of a Landau and a Gaussian function; the gain of the crystal is defined as the most probable value (MP) of the Landau function.  \label{fig:timecorr}}
\end{figure}

Figure~\ref{fig:timecorr} (left) shows the average time dependence of signals, obtained by a sample of clean events extracted from the muon run. 
The distribution was obtained by accumulating signals from several crystals, each one normalized by dividing for the most probable value (MP) of the MIP signal distribution (see right panel of figure~\ref{fig:timecorr} and  section~\ref{single:gain}). 
The data were interpolated with the function defined by Eq.~\eqref{eq:timecorr}, where $T_0$ was fixed to its nominal value, $T_R+T_H$ (see section~\ref{sec:chip}), 
and four parameters were left to vary: the normalization constant, the decay times  and the relative weight between the two. 
Hence, the attenuation function $a(\Delta t)$ was evaluated by dividing the fitted function by its value evaluated at $\Delta t = 0$, so that $a(0) = 1$. 
The major component has a decay time of about 1.8~$\mu$s, which is roughly consistent with the characteristic emission time of the CsI(Tl)~\cite{pdg}. 
The strong attenuation of the signal for values of $\Delta$t above 17~$\mu$s is instead consistent with an  additional slower component, which amount to about 6$\%$ of the total signal and has a decay time of about 11~$\mu$s. 
It has to be noticed that the time-profile fitting described above has the only purpose of correcting the measured scintillation signals.
A realistic physical interpretation of the observed time effect in terms of scintillation components, would require a detailed response model of the HIDRA chip, which has more sophisticated scheme and control sequence than that  described in section~\ref{sec:chip}, and the unfolding of the PD spectral response function.  This goes beyond the purpose of this work.

\subsubsection{LPD gain equalization}
\label{single:gain}

\begin{figure}[t]
\begin{center}
\includegraphics[width=7.5cm]{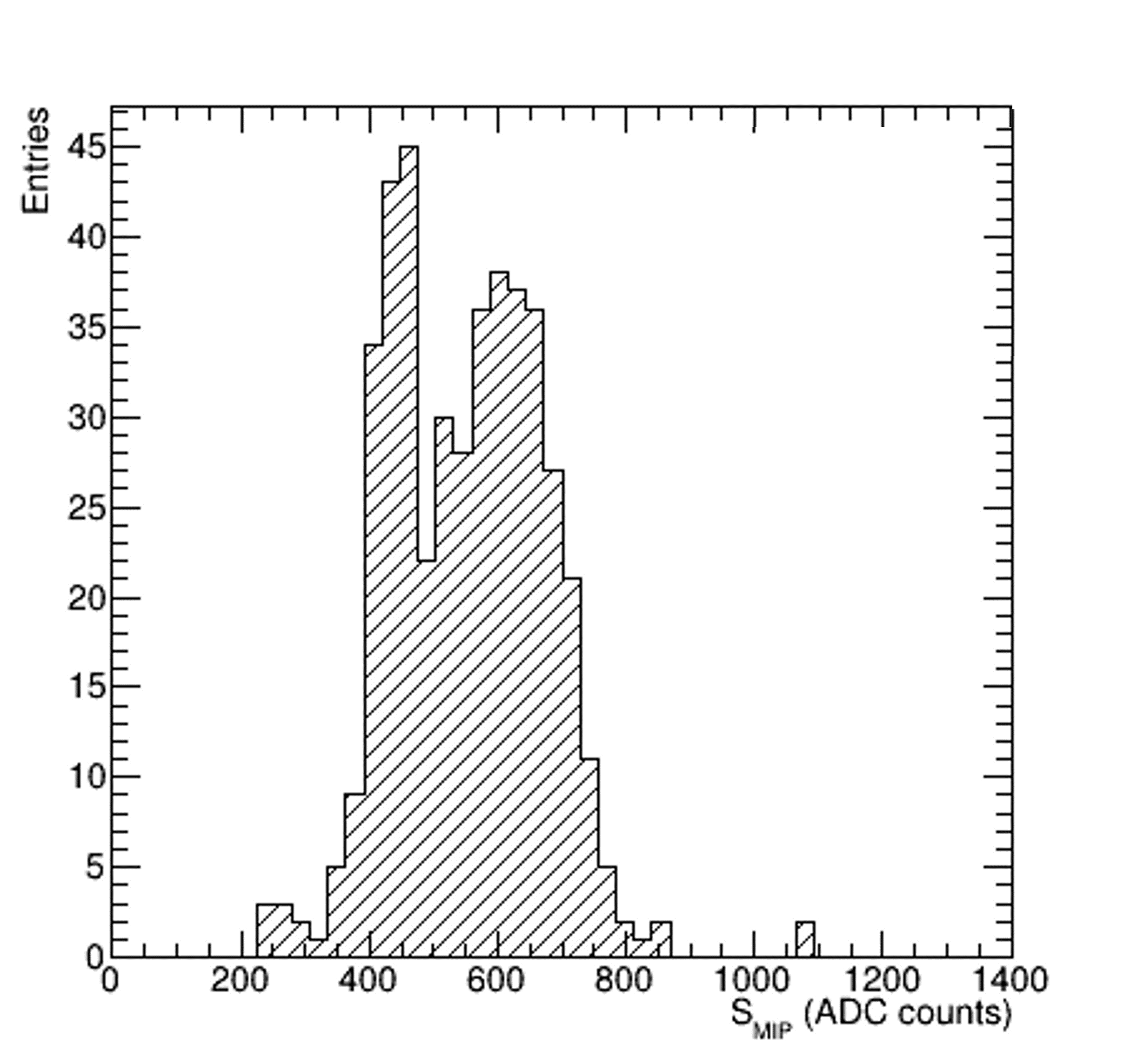} 
\end{center}
\caption{Distribution of the large-PD gain values ($S_{MIP}$) for all the crystals of the prototype.   \label{fig:mippa}}
\end{figure}

The typical distribution of signals generated by muons  corrected for the time attenuation, $S/a(\Delta t)$, is shown in figure~\ref{fig:timecorr} (right), for one of the LPDs.
The histogram is interpolated with a function defined as the convolution of a Landau and a Gaussian function. 
The line and the text panel in figure~\ref{fig:timecorr} (right) show the result of the fit for the considered LPD.  
One of the free parameters is the most probable value (MP) of the Landau distribution, which is the value $S_{MIP}$ 
used to normalize the response of each LPD channel. 

A full scan of all crystals was done, by collecting muon data with the calorimeter placed in different positions relative to the  beam axis. 
Figure~\ref{fig:mippa} shows the distribution of the normalization parameters obtained for all the crystals of the calorimeter. 
The average MIP signal is 560~ADC~counts, resulting in an overall MIP sigma-to-noise ratio for the calorimeter of  $\approx 7$. 

After the equalization, signals are 
expressed in arbitrary units that, by convention, are referred to as  MIP units through this paper.

\subsection{SPD gain calibration}
\label{sec:spdgain}

Each crystal is equipped with two PDs that measure the same scintillation light  with a different active area. 
The prototype system has been designed in such a way to have a reduced dynamic range, compared to what is required by the cosmic-ray mission. 
This design choice provided a wide overlapping  linearity region between the two PDs that 
allowed for a better cross-calibration of their gains and a more detailed study of the instrument systematic effects.

\subsubsection{SPD gain equalization}

\begin{figure}[t]
\begin{center}
\includegraphics[width=7.5cm]{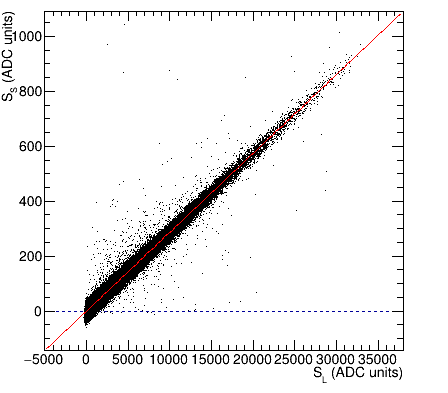}
\end{center}
\caption{Signal readout from the SPD  versus  signal readout from the LPD, for one of the calorimeter crystal. The red line represents a linear fit of the data. \label{fig:spdcalibration}}
\end{figure}

\begin{figure}[t]
\begin{center}
\includegraphics[width=7.5cm]{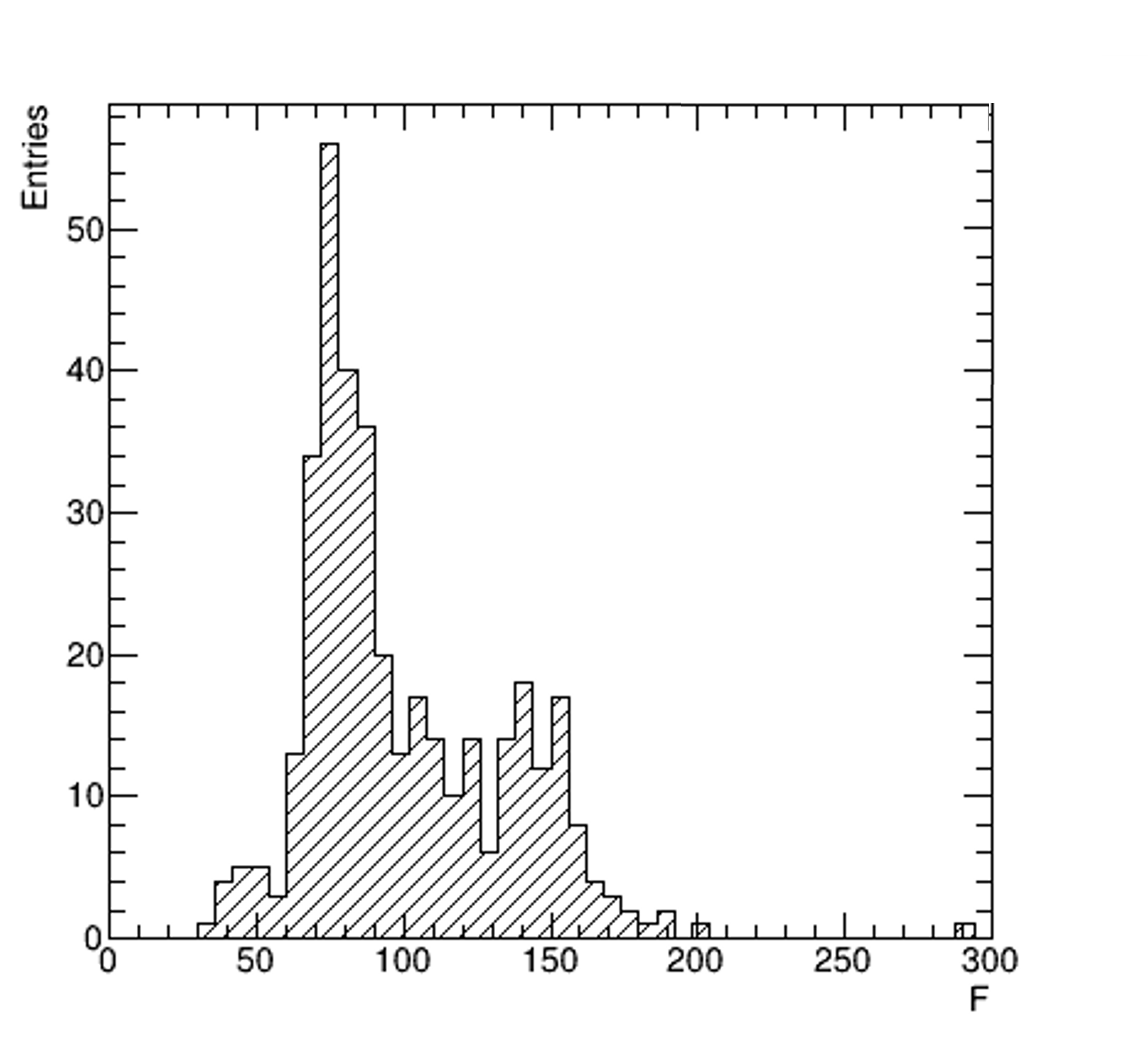}
\end{center}
\caption{ Distribution of the relative gain $F=S_{LPD}/S_{SPD}$ between the two photodiodes, large  and small, connected to each CsI(Tl) crystal. \label{fig:spdcalibrationset}}
\end{figure}

While the LPD gain can be determined by measuring the MP value of the MIP signals, the SPD gain is too low to detect it. 
Thus, the SPD was calibrated by evaluating its relative gain with respect to the LPD associated to the same crystal.
The single-crystal response to large energy deposits was studied by using data extracted from the hadron runs, which allow to test the crystal behaviour with energy deposits spreading out over the whole detector volume and the whole dynamic range. 
Fig~\ref{fig:spdcalibration} shows the correlation between the SPD and LPD signals 
from a crystal where the energy
deposit generates signals falling in the high-gain regime of the LPD. 
The  slope of the distribution depends on the relative gain ($F=S_L/S_S$) of the two PDs and was estimated, for each crystal, by performing a linear fit. 
The distribution of the measured parameters $F$ is shown in figure~\ref{fig:spdcalibrationset}, for all crystals that collect on average enough energy to generate a significant signal on the SPD. 
The relative gain between the large and the small PD is $\approx 100$, with a large variation among sensors that was attributed to the non-optimal optical coupling between the sensors and the crystals. 
The SPD signals were expressed in MIP units dividing their value by $F \cdot S_{MIP}$, after time attenuation correction~(section~\ref{single:time}).

\subsubsection{Calibration cross-check}

The SPD calibration parameters were cross checked by studying the signal correlation with the LPD signals acquired in low gain.
The  switch to low-gain mode occurs only for those crystals that collect the energy deposit from the shower cores. 
For this subset of crystals an independent estimate of the relative gain 
was derived 
and the parameters were compared with those obtained in high gain ($F$). 
A linear fit of the SPD-vs-LPD signals was done   
assuming the following function:
\begin{equation}
S_S = \frac{S_L + \Delta P}{F + \Delta F},
\label{eq:p0p1}
\end{equation}
where $\Delta P$ accounts for a possible residual pedestal shift of the low-gain LPD signal~(Eq.~\ref{eq:lg}), while $\Delta F$ expresses the difference of the SPD relative gain with the value derived in high-gain.
\begin{figure}[t]
\begin{center}
\includegraphics[width=7.5cm]{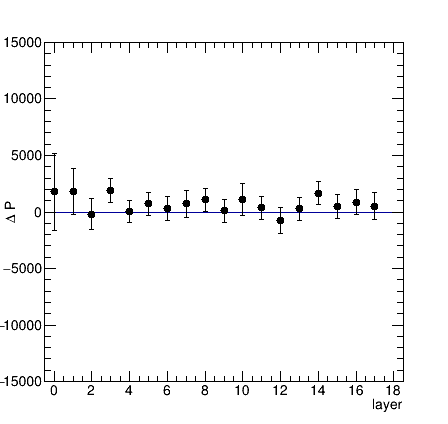}
\includegraphics[width=7.5cm]{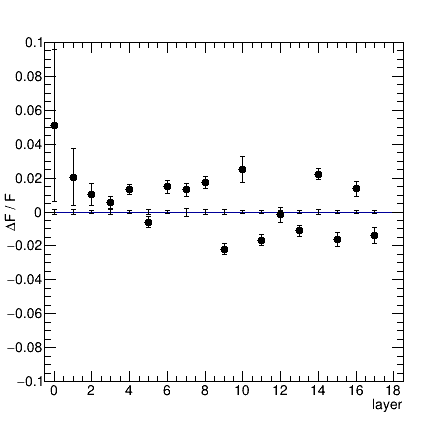}
\end{center}
\caption{  Residual pedestal shift (left panel) and relative SPD gain (right panel) obtained by comparing the SPD signal with low-gain LPD signal, for all the central crystals along the beam axis. 
In the right panel, the error bars along the horizontal axis of the plot show the uncertainty of the parameter $F$ determined with LPD signals in high gain. \label{fig:lgcheck}}
\end{figure}
The result of the low-gain linear fit is shown in Fig~\ref{fig:lgcheck} for all crystals along the beam axis. 
A small pedestal shift, $\Delta P \approx$~600~ADC counts (left panel), can be deduced from the data, which is however not quantitatively relevant for the shower reconstruction. 
The SPD relative gain (right panel) shows instead a significant discrepancy with the high-gain value,  $\Delta F/F \approx$ 2$\%$. 
A relative gain difference $\Delta F$ can be the result of multiple effects, either related to SPD response or to the low-gain LPD signal (Eq.~\ref{eq:lg}).  
A systematic pattern emerges from figure~\ref{fig:lgcheck} (right panel), which can be attributed both to the alternate position of the large and small sensors relative to the beam axis  and to the PD readout pattern, since crystals along the beam line are connected in pairs to the same FE chip by the $a$~and~$b$ sides of the same  cable (see figure~\ref{fig:pro} and \ref{fig:mec}). 
A cross-check of the calibration procedure done with simulated data (section~\ref{sec:simulation}) did not show any systematic effect in the determination of the relative SPD gains. 
As a whole, the observed effect (figure~\ref{fig:lgcheck}) is an indication of a systematic uncertainty in the determination of the absolute shower signal of at least $\approx$~2$\%$.  
%

\section{Instrument simulation}
\label{sec:simulation}

A detector simulation code was developed for the prototype, based on the FLUKA simulation tool~\cite{flukaa,flukab}. 
The full detector geometry was implemented, including CsI scintillating crystals, Si PDs and all relevant passive support structures (Polyoxymethylene trays). 
Mono-energetic particles of different types were generated, directed along the major axis of the calorimeter and with spatial coordinates distributed as the observed beam profiles. 

\subsection{Signal normalization}
For each channel $i$, the output of the simulation was expressed in ADC counts using the following relation:
\begin{equation}
    S_{i}^{sim} =  E_{i}^{ CsI} \cdot f_{i}^{ CsI} \cdot a(\Delta t) +    E_{i}^{ Si} \cdot f^{ Si} 
    + \Delta S_{i}^{ CsI} + \Delta S_{i}^{ Si} 
    + \Delta S_{i}^{ noise},
    \label{eq:sim}
\end{equation}
where  
$E_{i}^{CsI}$ and $E_{i}^{Si}$ are the energy deposits in the CsI crystal and in the depletion region of the PD (large or small), respectively; 
$f_{i}^{CsI}$ and $f_i^{Si}$ are the conversion factors used to scale from energy to ADC counts; 
$a(\Delta t)$ is the attenuation function of the signal defined in section~\ref{single:time}, with $\Delta t$  randomly generated event-by-event with a uniform distribution within the range $0 \div T_0$;
$\Delta S_{i}^{CsI}$ and $\Delta S_{i}^{Si}$ are random fluctuation terms that account for the stochastic processes of photon emission and collection, in CsI, and e-h pair creation, in Si; 
$\Delta S_{i}^{noise}$ is the detector noise.

The signal generated by the scintillation light depends on several processes, namely, the light yield of the crystal, the light collection efficiency and the responsivity of the PDs. 
According to the simulation, the energy deposit of 50~GeV muons in a $3.6$~cm tick CsI crystal is $\approx$~21.6~MeV, at the peak of the distribution defined by the MP value of a fitted Landau function.
In order to account for the real 
response of each channel, 
the energy deposit in the CsI crystals was normalized to the measured MIP signals, by setting the scaling factor of each channel $i$ to $f_{i}^{CsI} = S_{MIP,i} / 21.6$~MeV.
According to the results discussed in section~\ref{single:mip}~and~\ref{sec:spdgain}, this factor is on average $\approx$~25 ADC counts per MeV for the signals collected by the LPDs and $\approx$~100 times smaller for the SPDs. 
The average energy deposit of MIPs in 150~$\mu m$ of Si is about 37.4~keV. 
From previous studies~\cite{miriam}, an average prompt signal of $\approx$~60~ADC~counts was estimated for MIPs that vertically cross the LPD depletion region;  this value was used to derive the scaling factor for Silicon, $f_{Si} \approx$~1600~ADC counts per GeV.

From the scaling factor $f_{Si}$, assuming an average value of 1 e-h pair per 3.6~eV of deposited energy in Si, it results that one ADC count corresponds to about 200 charge units. 
According to this value, the measured MIP signal corresponds to an average number of photoelectrons generated on the LPD of the order of 10$^5$, which, given the nominal light yield of the CsI(Tl) and the QE of the LPD, is consistent with a light collection efficiency of $\approx$~10$\%$. 
The terms $\Delta S_{i}^{CsI}$ and $\Delta S_{i}^{Si}$ in Eq.~\ref{eq:sim}), were evaluated by converting the single energy deposits, in CsI and Si, from ADC counts to detected charge units, by smearing them according to a Poisson distribution and by converting them back to ADC counts.

The total signal, resulting from the sum of the CsI and Si energy deposits, expressed in ADC counts, was summed up to the channel measured pedestal and compared to the gain switch threshold to set the board gain and add the instrument noise. 
The electronic noise was simulated by randomly extracting off-spill events during data taking acquisition, after pedestal and common-noise subtraction.

Finally, the same reconstruction procedure (section~\ref{single:noise}) was applied to simulated and beam data. 

This normalization procedure allowed to reproduce the exact performances, in terms of signal-to-noise ratio, of the calorimeter prototype in the beam test configuration.

\begin{figure}[t]
\begin{center}
\includegraphics[width=7.5cm]{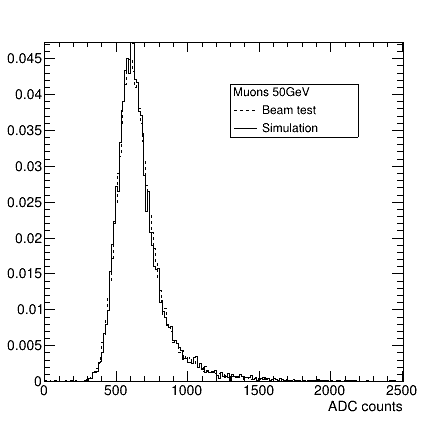}
\end{center}
\caption{Normalized distribution of the 50~GeV muon signal in a single crystal obtained from beam test data (solid line) and simulation (dashed line).  \label{fig:mcnorm}}
\label{fig:mipmc}
\end{figure} 
Figure~\ref{fig:mipmc} shows the single-crystal signal distribution obtained from 50~GeV simulated muons in comparison with experimental data.


\section{Response to electrons}
\label{sec:performance}

\begin{figure}[t]
\begin{center}
\end{center}
\hspace*{-0.4cm}
\includegraphics[width=16.1cm]{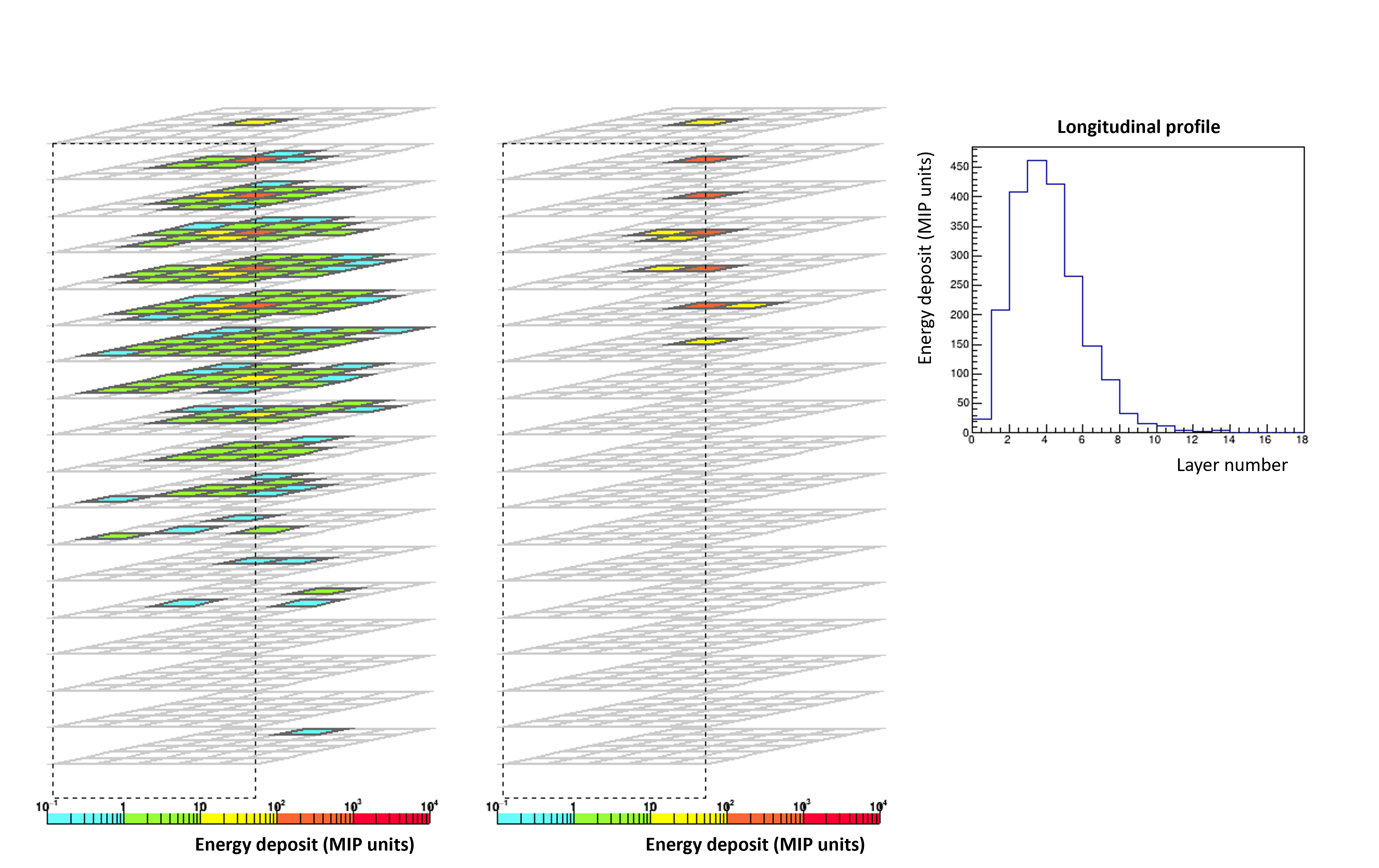}
\caption{ Event generated by a 50~GeV electron. The event image on the left~(right) panel represents the shower reconstructed by the LPDs~(SPDs); only crystals with a signal greater than~0.6~MIP~(14~MIP) are shown. The plot on the right shows the longitudinal shower profile, which in this case is done by using the LPD information only. }
\label{fig:evele50}
\end{figure}

\begin{figure}[t]
\begin{center}
\end{center}
\hspace*{-0.4cm}
\includegraphics[width=16.1cm]{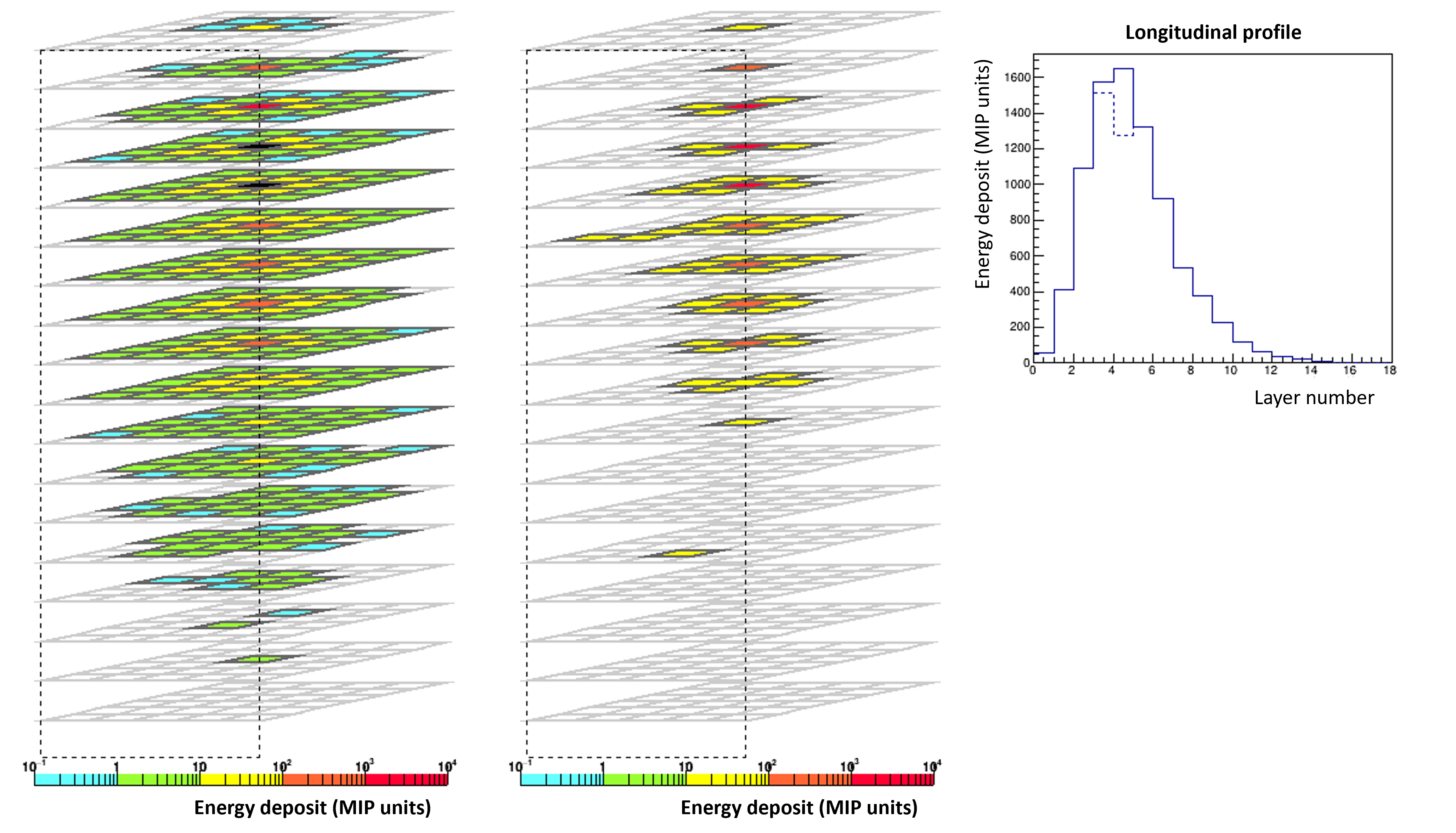}
\caption{Event generated by a 200~GeV electron. See figure~\ref{fig:evele50} for explanation. In this picture the black color indicates saturated signals. The plot on the right shows the reconstructed longitudinal profile of the shower, done by using the LPD information only (dashed line) and by using the SPD information when the LPD signal is above the saturation threshold  (solid line). }
\label{fig:evele300}
\end{figure}

The calorimeter response to electromagnetic showers represents a critical indicator of the instrument performances, since the intrinsic energy resolution is expected to be at the level of few $\%$. 
The prototype was exposed to electron beams of energy spanning from 50 to about 300 GeV.

\subsection{Shower reconstruction}

For each event, particle hits are identified by applying a cut of 0.6~MIP to the signal collected by the LPDs. 
A threshold value of 51000~ADC~counts in low gain was considered for all the raw signals in order to identify channel saturation. 
In case that the LPD signal was above the saturation threshold, the hit signal  was replaced with the SPD value. 
This happened at different amounts of energy deposit around the value of $\approx$~1200 MIP~units,  depending on the SPD gain.

Figure~\ref{fig:evele50} and \ref{fig:evele300} show two single events of  50~GeV and 200~GeV energy,  respectively.
The showers are shown both as seen independently by the LPDs and the SPDs and as reconstructed by combining the two. 

\subsection{Energy resolution}
\label{electrons:resolution}

\begin{figure}[t]
\begin{center}

\includegraphics[width=7.5cm]{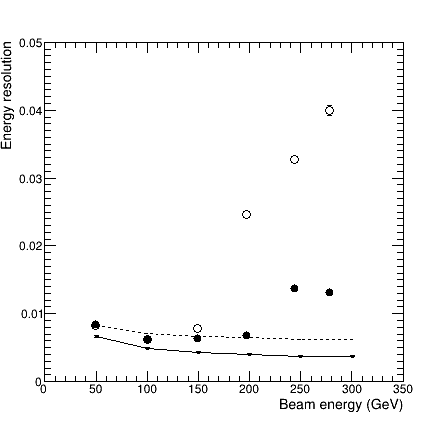}
\includegraphics[width=7.5cm]{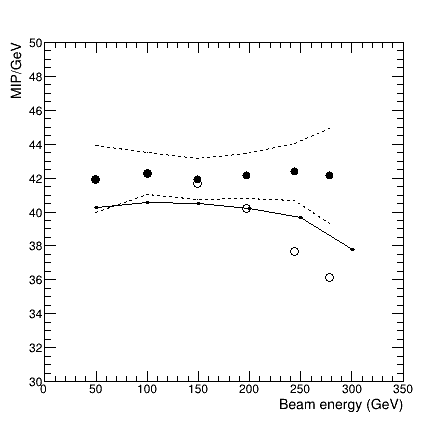}
\end{center}
\caption{ Measured energy resolution (left) and average total collected signal divided by the beam energy  (right) for electrons, as a function of the beam energy, obtained with the LPDs only (empty markers) and with the combined SPD and LPD signals (solid markers). 
The two plotted quantities were obtained from the Gaussian fit of the shower-signal distributions (see figure~\ref{fig:emdist}), as $\sigma/E_0$ and $E_0/E_{beam}$ respectively.   
The solid line shows the result of the simulation, for the combined LPD and SPD signals, in the ideal case of exact calibration parameters.
The dashed line on the left shows the result of a 0.5$\%$ additional fluctuation term added in quadrature to  the simulated energy resolution, while the dashed lines on the right show the  error band on the total collected signal obtained considering all the systematic uncertainties. 
See the text for more explanations. }
\label{fig:emres}
\end{figure}

\begin{figure}[p]
\begin{center}
\includegraphics[width=7.1cm]{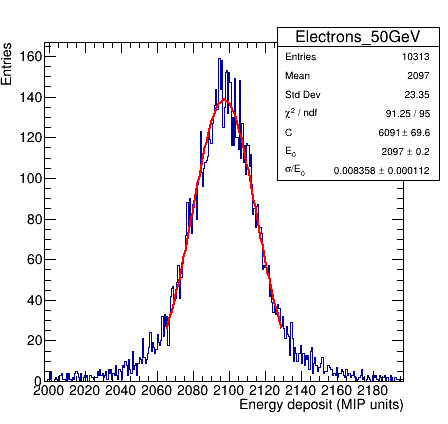}
\includegraphics[width=7.1cm]{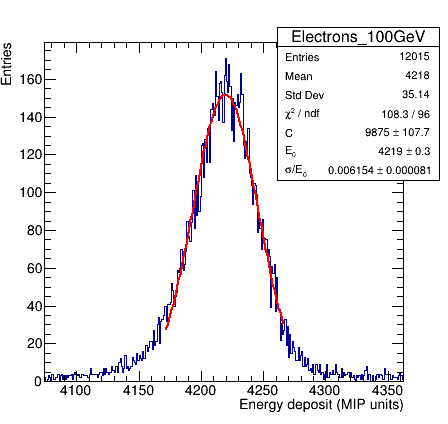}
\\
\includegraphics[width=7.1cm]{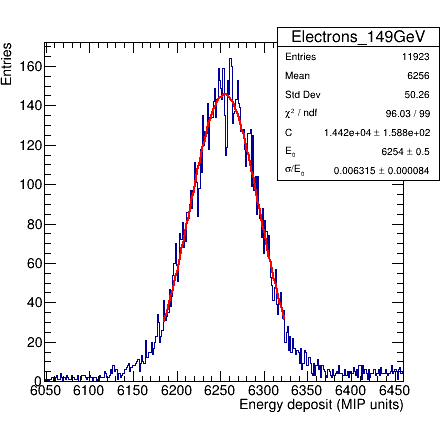}
\includegraphics[width=7.1cm]{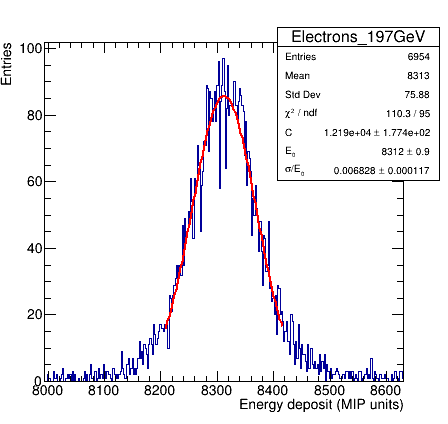}
\\
\includegraphics[width=7.1cm]{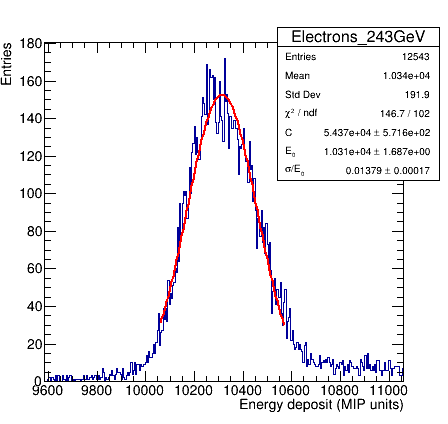}
\includegraphics[width=7.1cm]{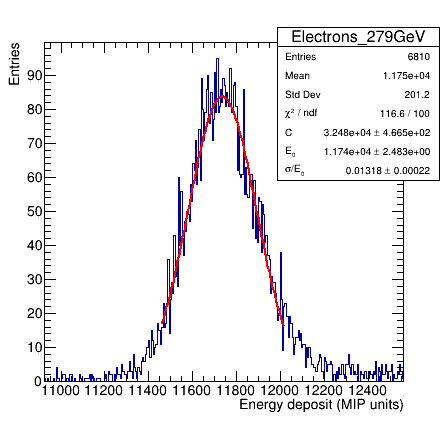}
\end{center}
\caption{ Distribution of the total collected signal, for electrons of different energies, obtained combining the SPD and LPD signals. The distributions were interpolated with a Gaussian function (red line) of width $\sigma$ and mean value $E_0$. }
\label{fig:emdist}
\end{figure}

\begin{figure}[t]
\begin{center}
\includegraphics[width=7.5cm]{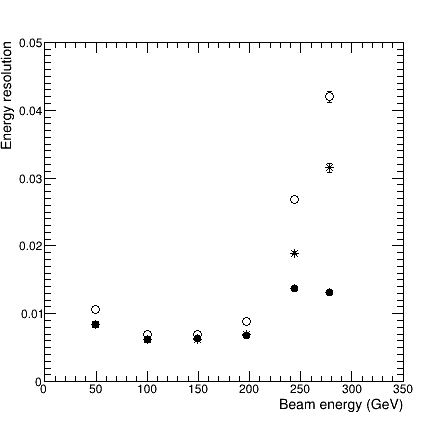}
\includegraphics[width=7.5cm]{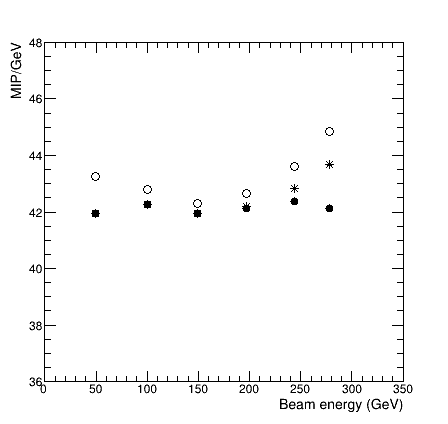}
\end{center}
\caption{Measured energy resolution (left) and average total collected signal divided by the beam energy (right) for electrons, as a function of the beam energy, obtained with the combined SPD and LPD signals, without common noise subtraction (empty markers), with standard (Eq.~\ref{eq:pedcnsub} and \ref{eq:lg}) common noise subtraction (asterisks) and  with saturation correction (solid markers). }
\label{fig:emrescn}
\end{figure}

The left panel of figure~\ref{fig:emres} shows 
the energy resolution,  obtained from the Gaussian fit of the distributions of the shower signals  (figure~\ref{fig:emdist}), 
as a function of the  beam energy for electrons hitting the calorimeter within 1~cm around the central axis. 
Empty markers indicate the values obtained with the LPDs only, while solid markers refer to combined SPD and LPD signals. 
Above 100~GeV a few LPD channels start to saturate, consequently the LPD energy resolution worsens;   
switching to SPD signals allows to partly  
recover the calorimeter performances.
The black dots connected by a solid line in figure~\ref{fig:emres} show the result of the simulation, for the combined LPD and SPD signals in the ideal case of exact calibration parameters.
Up to 200~GeV the measured resolution is consistent with the simulation, 
if an additional fluctuation term of $\approx$~0.5$\%$ is assumed (dashed line in figure~\ref{fig:emres}, left). 
This value is consistent with the calibration uncertainties, e.g. on the time attenuation correction~(see section~\ref{single:time}), which is expected to introduce  shower-by-shower fluctuations depending on the acquisition time $\Delta t$. 
Above 200~GeV the measured energy resolution significantly deviates from this expectation.  
This discrepancy was ascribed to 
the charge injection effect observed for this detector prototype during strong saturation of the LPD channels (see section~\ref{single:noise}).   
In this case, the implemented correction  compensates for most of the charge injection but does not allow to fully recover the nominal  calorimeter performances (see left panel of figure~\ref{fig:emrescn}). 
An upgrade of the cabling scheme to route the PD signal to the FE electronics is planned, which will reduce both the system noise and the amount of charge injected into the SPD channels. 

The right panel of figure~\ref{fig:emres} shows the average value of the total collected signal obtained form the Gaussian fit (figure~\ref{fig:emdist}) as a function of the beam energy. 
While the measured resolution is consistent with the expectation, up to 200 GeV, the total detected signal shows a  systematic discrepancy of $\approx$~5$\%$ with respect to the simulation. 
The dashed band drawn around the measured points represents an approximate estimate of the systematic uncertainty affecting  the total shower signal. 
This quantity was derived by propagating the single-crystal calibration uncertainties to the whole shower and by summing in quadrature an additional term representing an estimation of a possible residual shift caused by the CN drift; the latter quantity was conservatively assumed to be of the same amount of the common-noise subtraction itself (see right panel of figure~\ref{fig:emrescn}). 
The observed discrepancy between experimental data and simulation is of the order of the estimated systematic uncertainty.

\begin{figure}[t]
\begin{center}
\includegraphics[width=7.5cm]{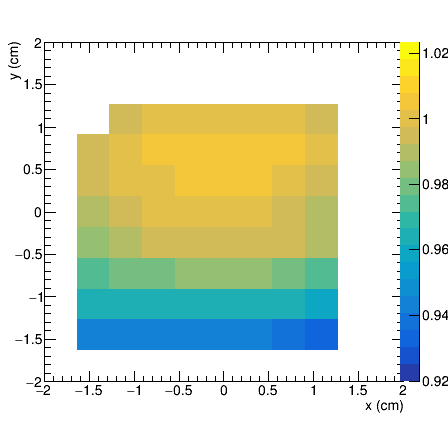}
\end{center}
\caption{  Relative variation of total collected signal as a function of the beam position, for normally incident electrons of 243~GeV. The origin of the coordinate system is placed at the center of the calorimeter.  }
\label{fig:emmap}
\end{figure}

The energy resolutions shown in figure~\ref{fig:emres} refer to electrons hitting the calorimeter within a narrow region of few cm$^2$ around the central axis. 
The total energy deposit is expected to vary with electron position and direction, at fixed energy, due to the presence of passive materials among the scintillating crystals and to the varying contribution of the direct ionization of PDs.  
In order to study the signal variation as a function of the particle position, a scan of the central crystals was done, by translating  the calorimeter  relative to beam line. 
Figure~\ref{fig:emmap} shows the relative variation of the total collected signal as a function of the particle position. 
The collected signal was found to be uniform at the level of about 2$\%$ for most of the scanned area, 
with larger discrepancies close to the crystal borders.
It has to be noticed that the beam-test configuration is not representative of the typical cosmic-ray detector exposure; in fact in terms of solid angle the fraction of events with incident direction close and parallel to the gaps among crystals is small and the overall effect on the energy resolution is expected to be significantly reduced~\cite{CC1}.

\section{Conclusions }

The final phase of the CaloCube project ended with the construction of a  large-scale prototype, 
consisting of 5$\times$5$\times$18 3.6~cm side cubic crystals made of CsI(Tl), for a total depth of $\approx$~27 radiation lengths. 
The prototype was tested with high-energy particle beams of different type. 
The content of the present paper focuses on the calibration of the instrument and on its  performance in response to high energy electrons.

The energy resolution for electromagnetic particles is a critical indicator of the instrument performance, since the expected energy resolution is of the order of few $\%$. 
A key issue for the next-generation cosmic-ray calorimetric experiments is the control of systematic uncertainties, as demonstrated by the most recent spectral measurements of the electron component, that are inconsistent with each other. 
A detailed study of the procedures to calibrate the prototype was done
and, thank to the redundancy of the system, an estimate of several possible sources of systematic uncertainty affecting the energy measurement was derived. 
The measured resolution ($\approx$~0.6$\%$ at 100~GeV for electrons normally hitting the calorimeter at the center) was found to be in good agreement with the expectation, if the calibration and instrumental uncertainties are considered. 

In spite of the excellent energy resolution, a systematic discrepancy of $\approx$~5$\%$ between the measured and simulated shower signal was found.
This discrepancy is marginally consistent with the estimated systematic uncertainties and might be an indication of either instrumental or physical effects  not accounted for in the instrument response model used in this work.

The tested prototype was characterized by a  not optimal design of the flat cables that route the PD signals to the FE electronics, which caused large noise and charge injection among nearby channels.
In spite of these undesired features, 
the design  allowed to effectively reduce their effect on the energy resolution and to restore the nominal performances up to 200~GeV of electron energy. 
Above this energy the charge injection induced on the SPD channels due to the saturation of the corresponding LPD channels  caused a degradation of the calorimeter performances. 

The results of this work guided the design of the later LYSO calorimeter prototypes, developed for the HERD experiment. 
Several changes were made to the design that are expected to improve the performance and reduce the systematic uncertainties.  
An improved version of the flat cable was developed, reducing the noise of the system and the capacitive coupling among nearby channels. 
A different cabling scheme to route the PD signals to the FE electronics was implemented; 
specifically, LPDs and SPDs were connected to separated FE chips through independent cables, so as to minimize the effect of any possible residual cross talk signal on the small SPD signals. 
Finally, being the LYSO response much faster than CsI(Tl), the time attenuation correction is expected to be negligible, which will reduce the calibration uncertainties and will increase the live time of the FE chip.

\bibliography{calocube}
\end{document}